\documentclass[11pt]{article}
\usepackage{latexsym}  
\usepackage{amssymb}
\usepackage{graphicx}
\usepackage{amsmath}

\begin{document}

\hspace*{11cm} {OU-HET-647/2009}

\begin{center}
{\Large\bf How Can $CP$ Violation in the Neutrino Sector}

{\Large\bf be Large in a $2\leftrightarrow 3$
Symmetric Model?}

\vspace{4mm}

\vspace{4mm}
{\bf Yoshio Koide$^a$ and Hiroyuki Nishiura$^b$}

${}^a$ {\it Department of Physics, Osaka University, 
Toyonaka, Osaka 560-0043, Japan} \\
{\it E-mail address: koide@het.phys.sci.osaka-u.ac.jp}

${}^b$ {\it Faculty of Information Science and Technology, 
Osaka Institute of Technology, 
Hirakata, Osaka 573-0196, Japan}\\
{\it E-mail address: nishiura@is.oit.ac.jp}

\date{\today}
\end{center}

\vspace{3mm}
\begin{abstract}
Based on a neutrino mass matrix model in which a 2-3 symmetry
is only broken by a phase parameter, it is investigated how
the lepton mixing matrix can deviate from the so-called
tribimaximal mixing under a condition that $CP$ is maximally
violated.
\end{abstract}

\vspace{3mm}

\noindent{\large\bf 1 \ Introduction}

Recent data in neutrino oscillation experiments 
\cite{atm,solar} have showed that the lepton mixing 
matrix $U$ is fairly in favor of the so-called tribimaximal 
mixing \cite{tribi} $U_{TB}$.
If $U$ is exactly $U=U_{TB}$, then $CP$ violation in the lepton sector
is absolutely forbidden.
Our interest is in how the $CP$ violation in the lepton
sector can be large and how $U$ can deviate from the 
tribimaximal mixing $U_{TB}$.

Meanwhile, we know that a neutrino mass matrix $M_\nu$ with
a $2\leftrightarrow 3$ symmetry \cite{23sym} can lead to 
a maximal $\nu_2$-$\nu_3$ mixing.
Of course, a rigorous requirement of the  $2\leftrightarrow 3$ 
symmetry for the neutrino and charged lepton sectors cannot 
give a realistic lepton mixing matrix $U$ because of SU(2)$_L$ 
symmetry \cite{K-T08}.
We have to bring a breaking term into the $2\leftrightarrow 3$ 
symmetric mass matrix by hand. 
A quark and lepton mass matrix model 
\cite{broken23sym} has been proposed, in which the 2-3 symmetry 
is broken only by a phase parameter:
A mass matrix $M_f$ is given by a form
$$
M_f = P(\delta_f) \widehat{M}_f P^\dagger(\delta_f) ,
\eqno(1.1)
$$
where
$$
P(\delta) = {\rm diag}(e^{i \delta_1}, e^{i\delta_2}, e^{i\delta_3}) ,
\eqno(1.2)
$$
$$
\widehat{M}_f = \left(
\begin{array}{ccc}
0 & a_f & a_f \\
a_f & b_f & -c_f \\
a_f & -c_f & b_f 
\end{array} \right) ,
\eqno(1.3)
$$
where $a_f$, $b_f$ and $c_f$ are real parameters.
[We have defined the expression (1.3) such as 
$m_{f3} >m_{f2} > -m_{f1} >0$ for $b_f >c_f>0$.
See Eq.(A.4) in Appendix.]
Here, the matrix $\widehat{M}_f $ is exactly symmetric under 
the $2\leftrightarrow 3$ flavor exchange, but the 2-3 symmetry 
for the mass matrix $M_f$ is broken by the phase factor 
$\delta_{f3}-\delta_{f2}\neq 0$. 
A short review of the diagonalization of the Herimitian mass 
matrix (1.1) is given in Appendix.

In this paper, we investigate $CP$ violation in the lepton 
sector.
Various models in which the 2-3 symmetry is broken only by 
phase parameters can be considered.
The most naive way is to consider a model 
$M_\nu = P(\delta_\nu) \widehat{M}_\nu P(\delta_\nu)$ 
similar to the form (1.1), and 
such a model has been discussed in Ref.\cite{broken23sym}. 
However, in the present paper, we take another idea that 
the $CP$ violation in the lepton sector is cased by phase 
parameters which cannot be removed by
a form $M_\nu = P(\delta_\nu) \widehat{M}_\nu P(\delta_\nu)$
differently from those in the quark sector.
The simplest realization of such a model is to assume 
a neutrino mass matrix form 
$$
M_\nu = \left(
\begin{array}{ccc}
0 & a_\nu & a_\nu \\
a_\nu & b_\nu e^{i(\phi+\chi)} & -c_\nu \\
a_\nu & -c_\nu & b_\nu e^{i(\phi-\chi)}
\end{array} \right) ,
\eqno(1.4)
$$ 
where $a_\nu$, $b_\nu$ and $c_\nu$ are real parameters.
Here, the 2-3 symmetry is broken only by the phase 
parameter $\chi$ as
$$
\frac{M_{\nu 22}-M_{\nu 33}}{M_{\nu 22}+M_{\nu 33}}
= i \tan\chi ,
\eqno(1.5)
$$
while $CP$ is broken by the phase
parameter $\phi$. 
We consider that the existence of these phase parameters is 
characteristic only in a Majorana mass matrix and
we cannot consider such the phases in  Dirac 
mass matrices (we assume that those are Herimitian).
The mass matrix can lead to a nearly tribimaximal mixing
when $CP$ is maximally broken in the lepton sector.
[For the charged lepton sector, we assume the matrix form (1.3).]

It is hard to solve the mass matrix model (1.4) analytically. 
Therefore, in Sec.2, we investigate the model (1.4) numerically.
The model (1.4) has four parameters $a \equiv a_\nu/b_\nu$, 
 $c\equiv c_\nu/b_\nu$, $\phi$ and $\chi$ for four observable 
quantities $\sin^2 2\theta_{atm}$, $\tan^2 \theta_{sol}$, 
$|U_{13}|^2$ and $R \equiv \Delta m^2_{sol}/\Delta m^2_{atm}$.
We will find that if we put a maximal $CP$ violation hypothesis, 
we can obtain interesting predictions.
Finally, Sec.3 is devoted to summary and concluding remarks.


\vspace{3mm}

\noindent{\large\bf 2 \ Lepton mixing matrix}

The lepton mixing matrix $U$ is given by
$$
U=U_e^\dagger U_\nu,
\eqno(2.1)
$$
where $U_e$ and $U_\nu$ are defined by
$$
U_e^\dagger M_e U_e = {\rm diag}(m_e, m_\mu, m_\tau) ,
\eqno(2.2)
$$
$$
U_\nu^\dagger M_\nu U_\nu^* = {\rm diag}(m_{\nu 1}, 
m_{\nu 2}, m_{\nu 3}) .
\eqno(2.3)
$$
A form of $U_e$ can readily be given by the   
analytic form $U_e=P(\delta_e) R_e$ given in Eq.(A.1).
In contrast to the case of $U_e$, it is hard to express 
$U_\nu$ with an analytic form. 
Therefore, in this paper, we discuss the parameter
dependences of the lepton mixing matrix $U$ by 
numerical studies.
Since, for $U_e$, we fix the mixing angle $\theta_e$ 
by inputting the charged lepton masses at $\mu=m_Z$,
the remaining free parameters for $U$ are $a_\nu/b_\nu$, 
$c_\nu/b_\nu$, $\phi$ and $\chi$.
(Since the value of $\theta_e$ is very small, i.e. 
$\theta_e=3.936^o$, the essential rotation in 
the lepton mixing matrix $U$ 
comes from the neutrino sector $U_\nu$.)
As stated in the previous section, the phase 
parameter $\chi$ breaks the $2\leftrightarrow 3$ 
symmetry, and $\phi$ causes the $CP$ violation.
In contrast to the phase parameters $\phi$ and $\chi$ 
in the lepton sector, there are two phase parameters 
$\alpha$ and $\beta$ in the quark sector, which are 
defined by Eq.(A.8).  
The parameter $\beta$ is only related to the breaking of 
the $2\leftrightarrow 3$ symmetry and $\alpha$ is 
only related to the $CP$ violation as shown in Eq.(A.9). 
However, note that those parameters in the quark sector 
do not affect 
quark mass values, while the phase parameters 
$\phi$ and $\chi$ affect not only the neutrino mixing 
matrix $U_\nu$ but also the neutrino mass values. 
In this paper, we will fix the phase parameters 
$\phi$ and $\chi$ by requiring a maximal $CP$ 
violation in the lepton sector.
The remaining parameters $a_\nu/b_\nu$ and $c_\nu/b_\nu$
can, in principle, be fixed by the observed values of
$$
R= \frac{\Delta m^2_{sol}}{\Delta m^2_{atm}} 
= \frac{m_{\nu 2}^2 -m_{\nu 1}^2}{m_{\nu 3}^2 
-m_{\nu 2}^2} ,
\eqno(2.4)
$$
(we define as $m_{\nu 2}^2 >m_{\nu 1}^2$) and 
$\tan^2 \theta_{sol} = {|U_{12}|^2}/{|U_{11}|^2}$.
Then, we can predict values of the rephasing invariant \cite{J}
$J={\rm Im}(U_{11} U_{22} U_{12}^* U_{21}^*)$, and the
neutrino oscillation parameters 
$\sin^2 2\theta_{atm}= 4|U_{23}|^2|U_{33}|^2$, 
and $|U_{13}|^2$.
However, the observed values of $R$ and 
$\tan^2\theta_{sol}$ are not so accurate, at present, 
to determine the parameter values $a=a_\nu/b_\nu$ and 
$c=c_\nu/b_\nu$.
Therefore, in this paper, we will show the predictions 
of $|J|$, $\sin^2 2\theta_{atm}$, $\tan^2\theta_{sol}$,  
$|U_{13}|^2$ and $R$ as a function of the parameters
$a$ and $c$ under 
the requirement of the maximal $CP$ violation.

We consider that the phenomenological mass matrix form (1.4) 
is given at an energy scale $\mu = M_Z$ ($M_Z$ is the 
$Z$-boson mass). 
Since the present model gives the neutrino mass hierarchy 
$m_{\nu 1}^2 \ll m_{\nu 2}^2 \ll
m_{\nu3}^2$ (not $m_{\nu 1}^2 \simeq m_{\nu 2}^2 \ll m_{\nu3}^2$), 
we can regard that the renormalization group equation effects are 
not so large for the mass ratios and mixing in the present model 
(see, for example, Ref.\cite{Matsuda04-evol}).  Therefore, 
the numerical calculation will be done at $\mu=M_Z$, 
though we suppose that a breaking of the 2-3 symmetry 
is caused at a TeV energy scale. 

Note that in the following subsections A-C, we sometimes show
figures with specific input parameter values.
However, those input values are temporary ones for convenience 
to demonstrate parameter dependence of the model, and
the values do not means final best-fit values in the present
analysis.
The final values are obtained from iteration of the analysis
A $\rightarrow$ B $\rightarrow$ C.

\vspace{2mm}

\centerline{\bf A. \ Behavior of $J$ versus phase parameters}

Our interest is in the maximal $CP$ violation.
The maximal $CP$ violation is defined as follow:
the rephasing invariant $J$ takes its maximal 
value for the $CP$ violating mass matrix parameters.
The $CP$ violating mass matrix parameter is a phase 
parameter $\alpha$ defined by Eq.(A.8) in the quark 
sector [the model (1.1)],
while it corresponds to the phase parameter $\phi$ in the 
present model (1.4).
We illustrate  $\phi$-dependence of $J$ for
typical values of $\chi$ in Figs.1.
Since $\phi \rightarrow -\phi$ (and 
$\chi \rightarrow -\chi$) means $U_\nu
\rightarrow U_\nu^*$, so that it means 
$U \rightarrow U^*$ because $U_e$ is real.
Note that cases with $(a, c)$ and 
$(-a, -c)$ give identical results for the 
observable quantities, while those give different
results for cases with $(a, -c)$ and $(-a, c)$. 
As far as the shape of $|J|$ versus $\phi$ is concerned,
the cases with $a c <0$ and $ a c >0$ are similar,
but the maximal values of $|J|$ are different from each
other as seen in Figs.1 (a) and (b). 
Hereafter, we will refer the cases with 
$a c <0$ and $ a c >0$ to as Cases A and B, 
respectively. 
As seen in Fig.1, the maximal $|J|$ takes place at 
$\phi \simeq  \pm 13^\circ$ for the cases 
$\chi=90^\circ$ and $\chi=120^\circ$ (and 
$\phi \simeq \pm 12^\circ$ for the case $\chi=60^\circ$)
for both cases A and B.  
(Thus, the maximum point is slightly dependent on the 
parameter $\chi$.) 
The case $\phi\simeq 13^\circ$ ($\phi \simeq 12^\circ$) 
gives a maximal $J$ with $J>0$, while $\phi \simeq -13^\circ$ 
($\phi \simeq -12^\circ$) 
gives a maximal $|J|$ with $J<0$. 
In this section, we take a standpoint that the maximal
$CP$ violation means a maximal $J$ ($J>0$).
Hereafter, we confine ourselves to investigating a parameter
region $0 \leq \phi \leq \pi/2$.
Especially, since our interest is in the maximal $CP$ violation,
we will investigate parameter dependences for the case 
$\chi=\pi/2$ in which the 2-3 symmetry is maximally broken.
On the other hand, in the limit of $\phi\rightarrow 0$,
a case $\chi \rightarrow \pi/2$ (and also $\chi \rightarrow -\pi/2$)
corresponds to a maximal 2-3 symmetry violation as seen in Eq.(1.5).
Therefore, we confine ourselves to investigating a parameter
region $0 \leq \chi \leq \pi/2$.

\vspace{2mm}

\centerline{\bf B. \ $(a, c)$ dependence}

Next, we investigate $(a,c)$-dependence of the observable  
values for $\sin^2 2\theta_{atm}$, $\tan^2 \theta_{sol}$, 
$|U_{13}|^2$, and $R$ in order to give more realistic 
predictions of them.
Since our concern is to predict the observable quantities
at a point in which $CP$ is maximally violated, we 
illustrate the $(a,c)$ dependence at $\chi=\pi/2$ and
$\phi =13^\circ$.
(The value of $\phi=13^\circ$ is not a final result, because 
the value of $\phi$  which gives a maximal $|J|$ in Fig.1 is 
dependent on the values of $a$ and $c$.)

Correspondingly to Cases A and B, $a$-dependences of the
predicted values are illustrated 
in Figs.2 (a) and (b), respectively, where the parameter 
$c$ is fixed at a typical value $c=0.95$ in both figures (a)
and (b).
We find that the present observed data \cite{atm,solar}
$$
\tan^2\theta_{sol} = 0.469^{+0.047}_{-0.041} \ \ 
(\theta_{sol}=34.4^{+1.3}_{-1.2} \ {\rm degrees})
\eqno(2.5)
$$
require $a=-(0.26^{+0.02}_{-0.03})$ and $a=+(0.29^{+0.04}_{-0.03})$
for Cases A and B, respectively, and 
$$
R_{obs} = \frac{(7.59 \pm 0.21)\times 10^{-5}\ {\rm eV}^2}{
(2.19^{+0.14}_{-0.13})\times 10^{-3}\ {\rm eV}^2}
= (3.47^{+0.43}_{-0.26}) \times 10^{-2} 
\eqno(2.6)
$$
require $a=-(0.26\pm 0.03)$ and $a=+(0.29^{+0.04}_{-0.03})$ for 
Cases A and B, respectively. 
From overall view, in Case A a value $a=-0.255$ can give favorable 
predictions of $\tan^2 \theta_{atm}$ and $R$,
so that we will use the the value $a=-0.255$ for Case A
hereafter.
For Case B, we will use a value $a=+0.270$ hereafter,
although the value does not so excellent agreement with
the observed values (2.5) and (2.6) as compared with
Case A. 

Also, $c$-dependence is illustrated in Fig.3 (a) and (b), 
which correspond to Cases A and B, respectively. 
Here, the parameter $a$ is fixed at $a=-0.255$ for
Case A (and $a=+0.270$ for Case B) from the results
in Fig.2.
We find that the predicted values of $\tan^2\theta_{sol}$
and $R$ are not so sensitive to the value of $c$. 
If we use a parameter value $c=0.95$ for both Cases A
and B, which is favorable to the data, we obtain predictions
$(\tan^2\theta_{sol}, R)=(0.468,0.0347)$ and $(0.443,0.0350)$
correspondingly to the input values 
$(a=-0.255,c=0.95)$ and $(a=+0.270,c=0.95)$.

\vspace{2mm}

\centerline{\bf C. \ Phase parameter dependence }

Again, in Fig.4, we illustrate behaviors of the predicted 
values $J$ versus the phase parameter $\phi$, 
together with the behaviors of  
$\sin^2 2\theta_{atm}$, $\tan^2 \theta_{sol}$, 
$|U_{13}|^2$ and $R$ under the maximal 2-3 symmetry
violation $\chi=\pi/2$. 
Here, the parameter values  $(a,c)=(-0.255,0.95)$ and 
$(a,c)=(+0.270,0.95)$ are taken in the figures (a) and
(b), respectively, from the study in the previous 
subsection.

From Fig.4, we find that the values of $\sin^2 2\theta_{atm}$
and $|U_{13}|^2$ are insensitive to the value of $\phi$, 
while the predicted values of $\tan^2 \theta_{sol}$ and $R$ 
are highly sensitive to the value of $\phi$. 
Exactly speaking, in Fig.4, since the parameters $a$ and $c$
have been taken so that the values $\tan^2 \theta_{sol}$ 
and $R$ are reasonably fitted, the values of $\tan^2 \theta_{sol}$ 
and $R$ are not predictions in Fig.4.
Our predictions are for $\sin^2 2\theta_{atm}$ and 
$|U_{13}|^2$: $\sin^2 2\theta_{atm} \simeq 0.98$ and 
$|U_{13}|^2 \simeq 0.007$ for Case A, and 
$\sin^2 2\theta_{atm} \simeq 0.93$ and 
$|U_{13}|^2 \simeq 0.035$ for Case B.

\vspace{2mm}

\centerline{\bf D. \ Predictions under the maximal $CP$
violation hypothesis}

It is interesting to assume a hypothesis of the maximal 
$CP$ violation.
As seen in Fig.4, although the value of $J_{max}$, 
$J_{max} \simeq 0.039$, in Case A 
is larger than $J_{max}\simeq 0.018$ in Case B, it does not 
mean that Case A is ruled out under this hypothesis.
The maximal $J$ is required only for the phase parameters,
not for the mass matrix parameters $a_\nu$, $b_\nu$ and 
$c_\nu$.
Therefore, both Cases A and B are allowed under the maximal
$CP$ violation hypothesis.
If we require a larger value of $J$, we have to choose Case B.
However, the Case B predicts $\sin^2 2\theta_{atm} \leq 0.93$, 
which contradicts the present atmospheric 
neutrino data \cite{Suzuki09} $\sin^2 2\theta_{atm} \geq 0.96$.
Therefore, we conclude that Case B should be ruled out, so 
that the mass matrix parameters $a_\nu$ and 
$c_\nu$ must have the opposite sign to each other.
In contrast to this conclusion, note that a sign of the 
parameter $a_f$ in a model (1.1) can be taken freely 
 as seen in Eqs.~(A.1) - (A.4).

In Fig.5, we illustrate the more detailed behavior of the 
physical observable quantities under the maximal $CP$ 
violation hypothesis.
Here, although Figs.5 (a) and (b) are similar to Fig.2 (a) and 
Fig.3 (b), respectively, Figs.5 (a) and (b) are illustrated 
under the maximal $CP$ violation hypothesis, so that the value 
$\phi$ is always taken as it gives the maximal $J$ for each 
parameter set of $a$ and $c$.
From Fig.5, we conclude that $\sin^2 2\theta_{atm} \simeq 0.98$
and $|U_{13}|^2 \simeq 0.007$ for the present data (2.5) and (2.6).
In Fig.6, we also illustrate the contour plots for $\sin^2 2\theta_{atm}$ 
(Fig.(a)) , $|U_{13}|^2 \times 10$ (Fig.(b)), and $J$  (Fig.(c)) 
in addition to $R\times 10$ and $\tan^2 \theta_{sol}$ given in 
Eqs.~(2.5) and (2.6) in the $(a,c)$ parameter plane under 
the maximal $CP$ violation hypothesis for each parameter set 
of $a$ and $c$.

\vspace{3mm}

\noindent{\large\bf 4 \ Summary}

In conclusion, we have assumed a neutrino mass matrix in
which a 2-3 flavor symmetry is violated only by a phase 
parameter, and which can lead to a nearly tribimaximal
neutrino mixing, and thereby, we have investigated behaviors 
of the neutrino oscillation parameters under the maximal $CP$ 
violation hypothesis. 
We find that the predicted values of $\sin^2 2\theta_{atm}$ and
$|U_{13}|^2$ are insensitive to the input values $R$ and 
$\tan^2\theta_{sol}$ for $R\simeq 0.035$ and 
$\tan^2\theta_{sol}\simeq 0.5$:
$$
\sin^2 2\theta_{atm} \simeq 0.98 , \ \ \ |U_{13}|^2 \simeq 0.01.
\eqno(4.1)
$$
Especially, we are interested in the case with $|R| = 0.0321$ which
is a lower bound of the observed value of $|R|$ as shown in Eq.(2.6), 
because the value gives us a upper bound of the predicted value for
$\sin^2 2\theta_{atm}$ and a lower bound of the predicted 
value for $|U_{13}|^2$.
As seen in Fig.6, our model rules out the following values 
$$
\sin^2 2\theta_{atm} > 0.989 , \ \ \ |U_{13}|^2 < 0.0046,
\eqno(4.2)
$$
if neutrino oscillation data establish $|R| >0.321$.
On the other hand, if the present upper bound of $|R|$,  
$|R|=0.0390$, is established, the following regions 
$$
\sin^2 2\theta_{atm} < 0.978 , \ \ \ |U_{13}|^2 > 0.0098 ,
\eqno(4.3)
$$
are also ruled out.
Those bounds (4.2) and (4.3) are within reach of near 
future experiments.
Therefore, if future experiments give us values in the 
regions (4.2) and (4.3), the present model will be ruled out.
On the contrary, if future experiments report observed 
values within 
$$
0.989 \geq \sin^2 2\theta_{atm} \geq 0.978 , \ \ \ 
0.0046 \leq |U_{13}|^2 \leq 0.0098 ,
$$
$$
0.021 \geq J \geq  0.013 , 
\eqno(4.4)
$$
correspondingly to $0.0321 \leq |R| \leq 0.0390$ and
$0.428 \leq \tan^2 \theta_{sol} \leq 0.516$, 
the neutrino mass matrix form (1.4) will become a promising
candidate of the neutrino mass matrix form.

Also, note that the predicted value of $\tan^2\theta_{sol}$
is highly correlated to the predicted value of $R$ under 
the values (4.1).  
This result is shown in Fig.7 by using only the observable 
quantities with fixing typical values for $\sin^2 2\theta_{atm}$  
such as $\sin^2 2\theta_{atm}=0.980$ and $0.985$.
This figure is also useful as a touchstone of the present model.

The neutrino mass matrix (1.4) is a very simple form and 
it has phase parameters which are characteristic only 
in a Majorana mass matrix.
We expect that the results will be checked in the very near 
future experiments.

\vspace{3mm}

\centerline{\large\bf Acknowledgment}

One of the authors (YK) is supported by the Grant-in-Aid for
Scientific Research, JSPS (No.21540266).

\vspace{4mm}

\centerline{\large\bf Appendix}

We give a short review of the diagonalization of the mass 
matrix (1.1) and the Cabibbo-Kobayashi-Maskawa (CKM) quark 
mixing matrix as an example of the flavor mixing matrix.
(Although we have denoted as $(M_f)_{11}=0$ for simplicity, it
is not essential for the following discussions. 
For the case  $(M_f)_{11}\neq 0$ in Eq.(1.1), 
it is only necessary 
that the parameter values $b_f$ and $\delta_{f3}-\delta_{f2}$ 
are redefined under a common shift of the mass eigenvalues 
$m_{fi} \rightarrow m_{fi}+ (\widehat{M}_f)_{11}$.)

The matrix $\widehat{M}_f$ given in Eq.(1.3) is diagonalized 
by a rotation
$$
R_f =R_1(-\frac{\pi}{4}) R_3 (\theta_f)
=\left(
\begin{array}{ccc}
\cos \theta_f & \sin \theta_f & 0 \\
-\frac{1}{\sqrt{2}}\sin \theta_f & 
\frac{1}{\sqrt{2}}\cos \theta_f & -\frac{1}{\sqrt{2}} \\
-\frac{1}{\sqrt{2}}\sin \theta_f & 
\frac{1}{\sqrt{2}}\cos \theta_f & \frac{1}{\sqrt{2}} \\
\end{array}
\right) \ ,
\eqno(A.1)
$$
as 
$$
R_f^T \widehat{M}_f R_f =D_f\equiv {\rm diag} (m_{f1}, m_{f2}, m_{f3}) ,
\eqno(A.2)
$$
where
$$
\cos\theta_f = \sqrt{\frac{m_{f2}}{m_{f2}-m_{f1}}} \ ,  \ \ \ 
 \sin\theta_f =\sqrt{\frac{-m_{f1}}{m_{f2}-m_{f1}}} \ ,
\eqno(A.3)
$$ 
$$
\begin{array}{l}
m_{f1}=
\frac{1}{2}
\left(b_f -c_f-\sqrt{8a_f^2 + (b_f-c_f)^2}
\right)  , \\
m_{f2}=\frac{1}{2}
\left(b_f -c_f+\sqrt{8a_f^2 + (b_f-c_f)^2}
\right)  , \\
m_{f3}= b_f +c_f.
\end{array}
\eqno(A.4)
$$
Here and hereafter, we use the following notations for the 
rotation matrices
$$
R_1(\theta) = \left(
\begin{array}{ccc}
1 & 0 & 0 \\
0 & c & s  \\
0 & -s & c 
\end{array}
\right) , \ \ \ 
R_2(\theta)= \left(
\begin{array}{ccc}
c & 0 & s  \\
0 & 1 & 0 \\
-s & 0 & c \\
\end{array}
\right) , \ \ \ 
R_3(\theta)= \left(
\begin{array}{ccc}
c & s & 0 \\
-s & c & 0 \\
0 & 0 & 1 \\
\end{array}
\right) ,
\eqno(A.5)
$$
where $c=\cos\theta$ and $s=\sin\theta$.
When we apply this model to quark mass matrices, we obtain
the CKM quark mixing matrix $V$
as follows:
$$
V=U_u^\dagger U_d = R_u^T P(\delta) R_d =
R_3^T(\theta_u) R_1^T(-\frac{1}{4}\pi) P(\delta) 
R_1(-\frac{1}{4}\pi) R_3(\theta_d),
\eqno(A.6)
$$
where $\delta_i=\delta_{ui} -\delta_{di}$ ($i=1,2,3$) and
$\tan\theta_f = \sqrt{-m_{f1}/m_{f2}}$.
We can rewrite the factor $R_1^T(-\frac{1}{4}\pi) P(\delta) 
R_1(-\frac{1}{4}\pi)$ as
$$
R_1^T(-\frac{1}{4}\pi) P(\delta) R_1(-\frac{1}{4}\pi) 
= \left(
\begin{array}{ccc}
e^{i\delta_1} & 0 & 0 \\
0 & \frac{1}{2} \left(e^{i\delta_2} +e^{i\delta_3}\right) &
\frac{1}{2} \left(-e^{i\delta_2} +e^{i\delta_3}\right) \\
0 & \frac{1}{2} \left(-e^{i\delta_2} +e^{i\delta_3}\right) &
\frac{1}{2} \left(e^{i\delta_2} +e^{i\delta_3}\right) 
\end{array} \right)
$$
$$
= e^{i(\alpha +\delta_1)} \left(
\begin{array}{ccc}
e^{-i\alpha} & 0 & 0 \\
0 & \cos\beta & i\sin\beta \\
0 & i\sin\beta & \cos\beta 
\end{array} \right) 
=e^{i(\alpha +\delta_1)} P_1(-\alpha) P_3(-\frac{1}{2}\pi)
R_1(\beta) P_3(\frac{1}{2}\pi) ,
\eqno(A.7)
$$
where
$$
\alpha = \frac{1}{2}(\delta_3+\delta_2) -\delta_1 , 
\ \ \ \beta=\frac{1}{2}(\delta_3-\delta_2) ,
\eqno(A.8)
$$
$P_1(\delta)={\rm diag}(e^{i\delta}, 1, 1)$ and
$P_3(\delta)={\rm diag}(1,1,e^{i\delta})$.
Since $P_3$ and $R_3$ are commutable each other, the phase
matrices $P_3(-\pi/2)$ and $P_3(\pi/2)$ can be eliminated 
by redefining phases of the up- and down-quarks, respectively, 
so that we can obtain a CKM matrix form as follows:
$$
V=R_3^T(\theta_u) P_1(-\alpha) R_1 (\beta) R_3(\theta_d) .
\eqno(A.9)
$$
This expression (A.9) is well-known as the Fritzsch-Xing 
expression \cite{CKM-FX} of the CKM matrix.
Note that the phase parameter $\alpha$  still plays a role as
a $CP$ violating phase, while the parameter $\beta$ plays 
only a role as a rotation parameter $\theta_{23}$.
It is known that the model is in favor of the observed 
values of the CKM parameters under a hypothesis  that
the $CP$ violation is maximally violated, i.e. $\alpha=\pi/2$.

\vspace{4mm}

\newpage

{\scalebox{1.0}{\includegraphics{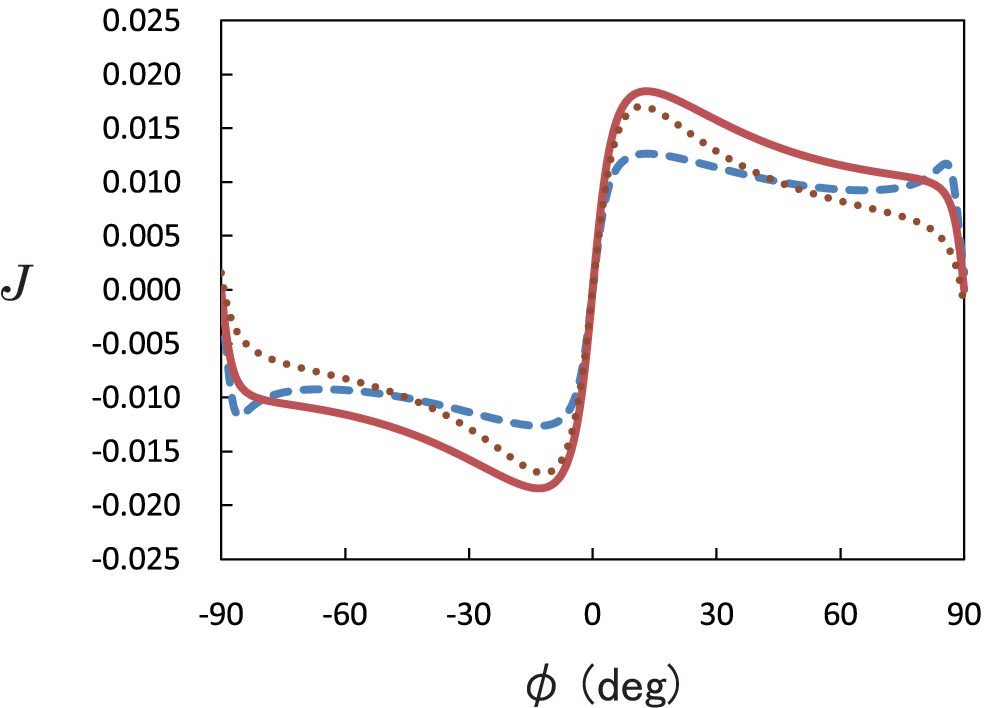}} }

\hspace*{3cm} {\bf (a)}

{\scalebox{1.0}{\includegraphics{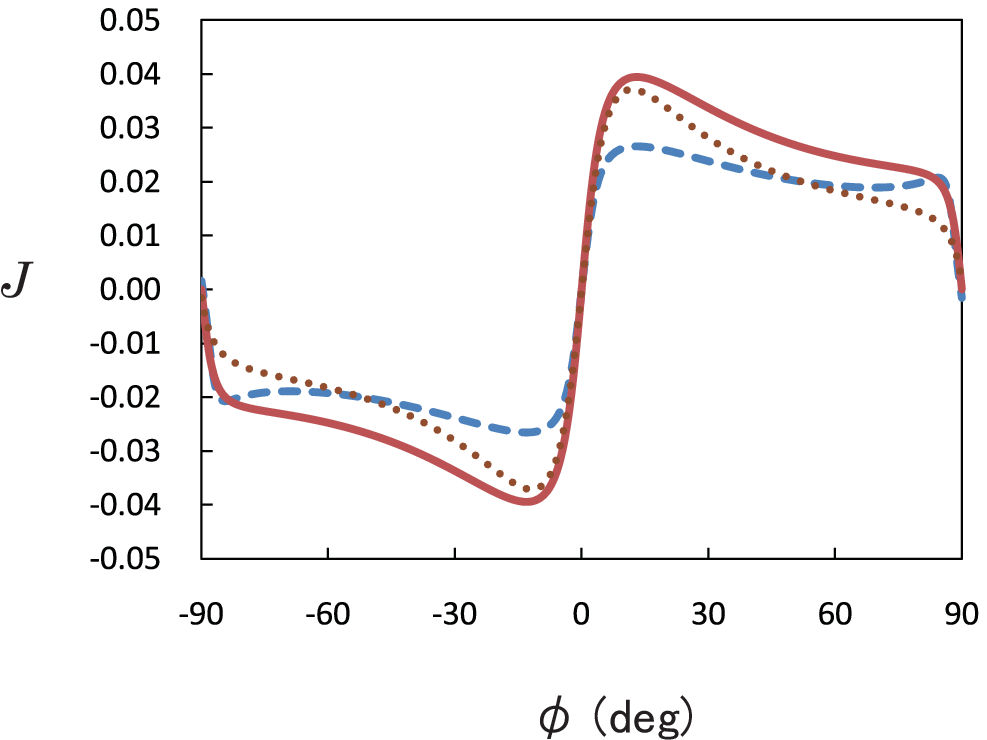}} }

\hspace*{3cm} {\bf (b)}

\begin{quotation}
{\bf Fig.~1}  
Behavior of the rephasing invariant $J$
versus the $CP$ violation parameter $\phi$. 
Figures (a) and (b) are illustrated for typical
values $(a,c)=(-0.27,+0.95)$ and 
$(a,c)=(+0.27,+0.95)$, respectively.
The dashed, solid, and dotted curves correspond to
cases with $\chi=120^\circ$, $\chi=90^\circ$,
and $\chi=60^\circ$, respectively.
\end{quotation}

\newpage

{\scalebox{1.0}{\includegraphics{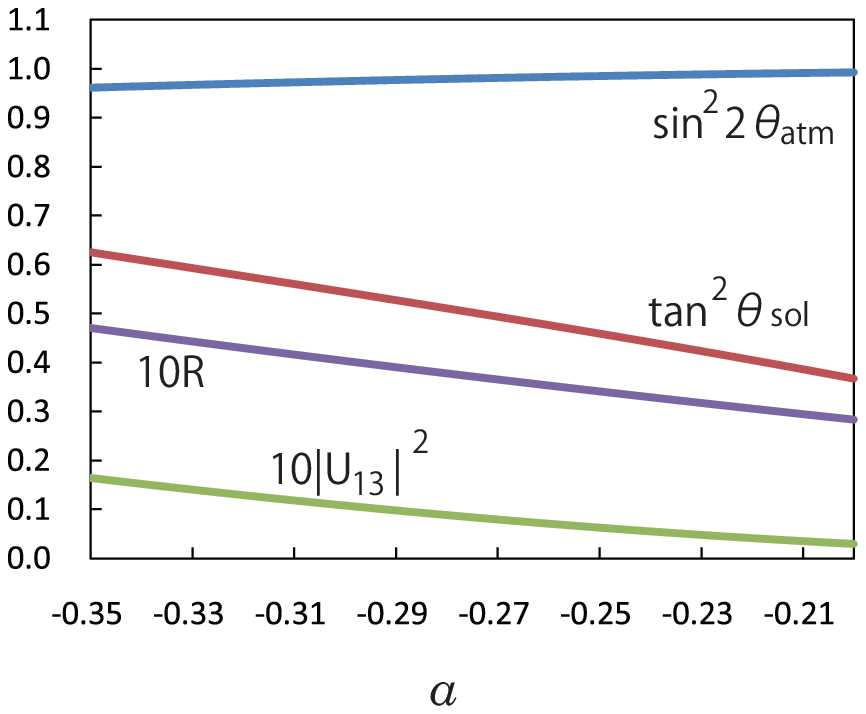}} }

\hspace*{3cm} {\bf (a)}

{\scalebox{1.0}{\includegraphics{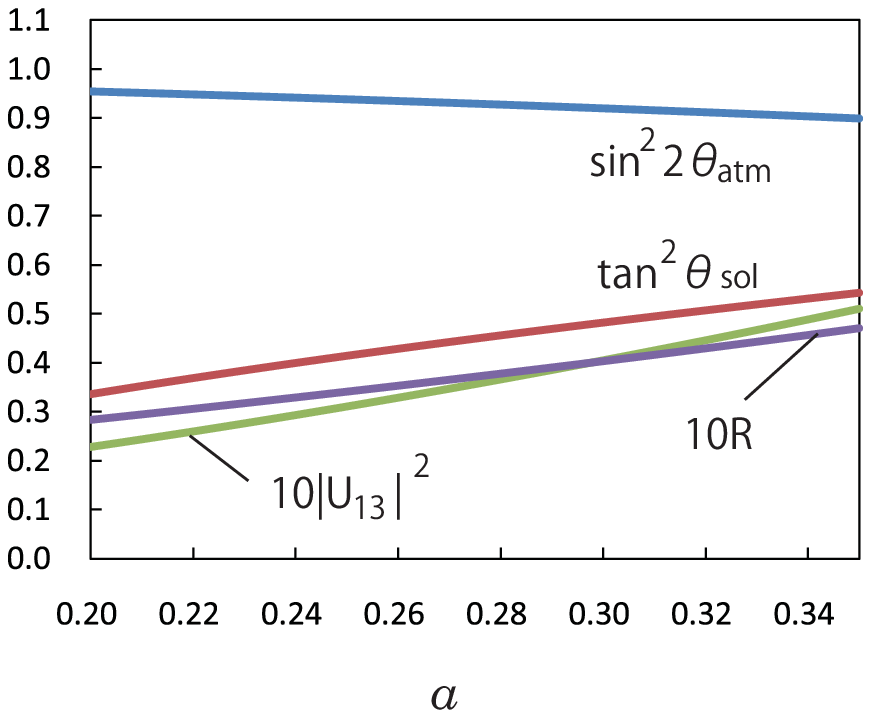}} }

\hspace*{3cm} {\bf (b)}

\begin{quotation}
{\bf Fig.~2}  
$a$-dependence of the predicted values for $\sin^2 2\theta_{atm}$, 
$\tan^2 \theta_{sol}$, $R\times 10$ and $|U_{13}|^2 \times 10$. 
Other parameters are fixed at typical values 
$c=c_\nu/b_\nu =0.95$, $\chi=\pi/2$, and $\phi=13^\circ$.
Figures (a) and (b) correspond to Case A with $a c <0$ and
Case B with $a c >0$, respectively.
\end{quotation}

\newpage

{\scalebox{1.0}{\includegraphics{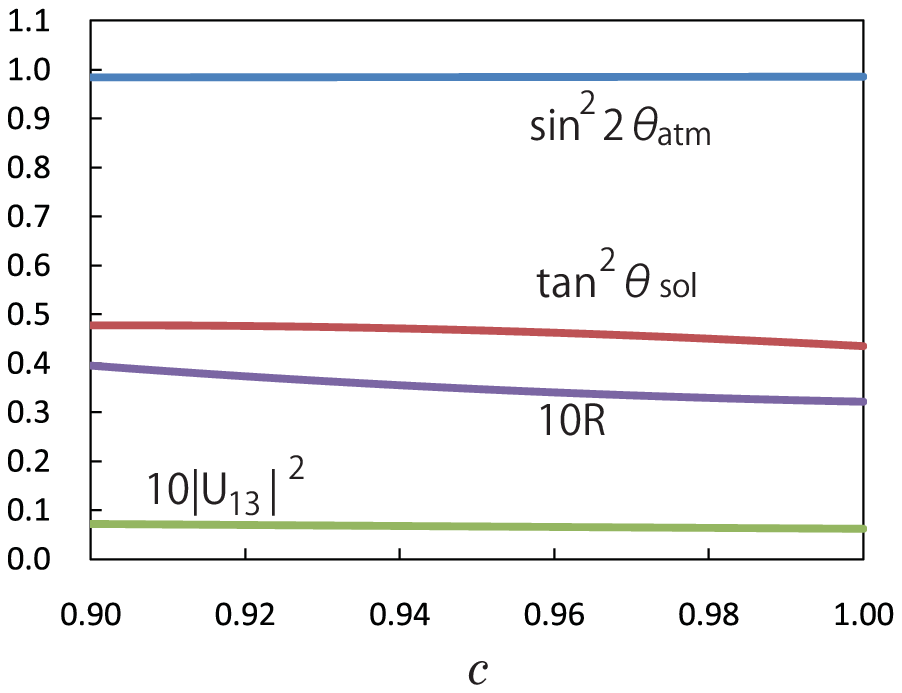}} }

\hspace*{3cm} {\bf (a)}

{\scalebox{1.0}{\includegraphics{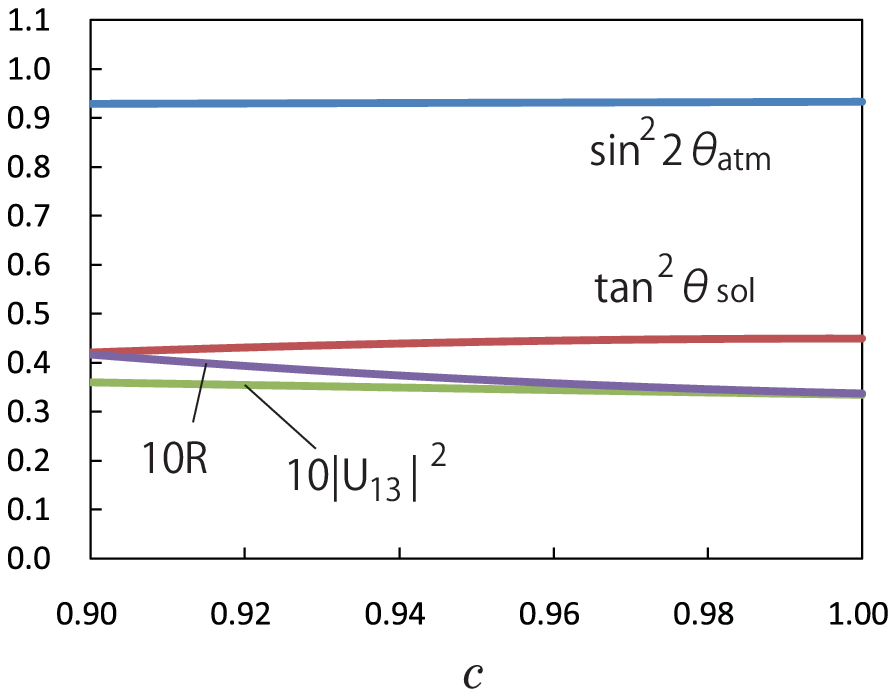}} }

\hspace*{3cm} {\bf (b)}

\begin{quotation}
{\bf Fig.~3}  
$c$-depencence of the predicted values for $\sin^2 2\theta_{atm}$, 
$\tan^2 \theta_{sol}$, $R \times 10$, and $|U_{13}|^2 \times 10$.  
Figures (a) and (b) correspond to Case A with $a c >0$ and
Case B with $a c <0$, respectively and those are illustrated
by taking input values $a=-0.255$ and $a=+0.27$, respectively.
Other parameters are fixed at typical values 
$\chi=\pi/2$ and $\phi=13^\circ$.
\end{quotation}

\newpage

{\scalebox{1.0}{\includegraphics{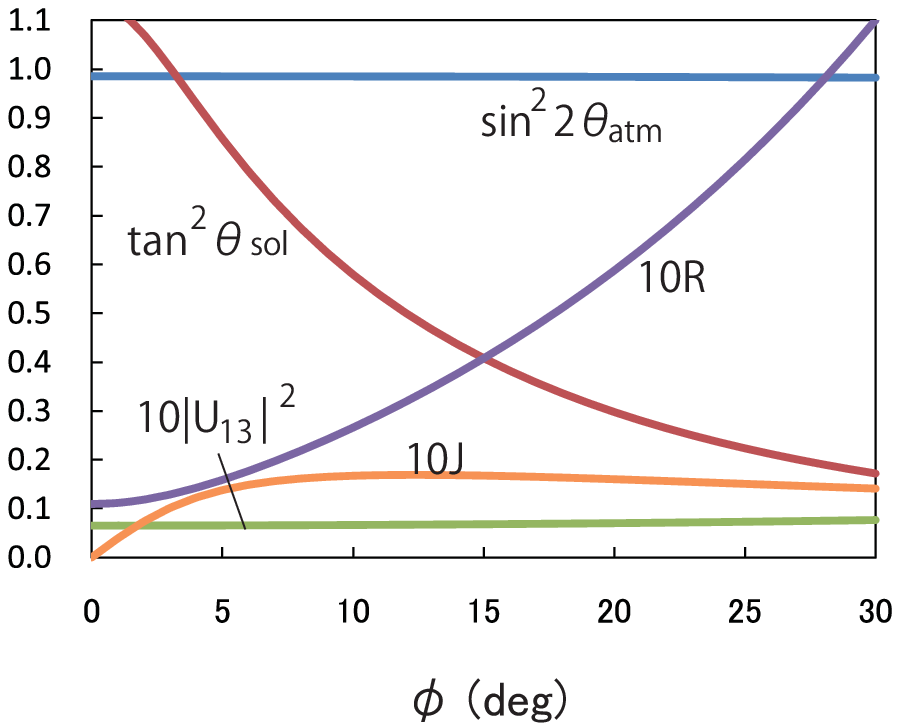}} }

\hspace*{3cm} {\bf (a)}

{\scalebox{1.0}{\includegraphics{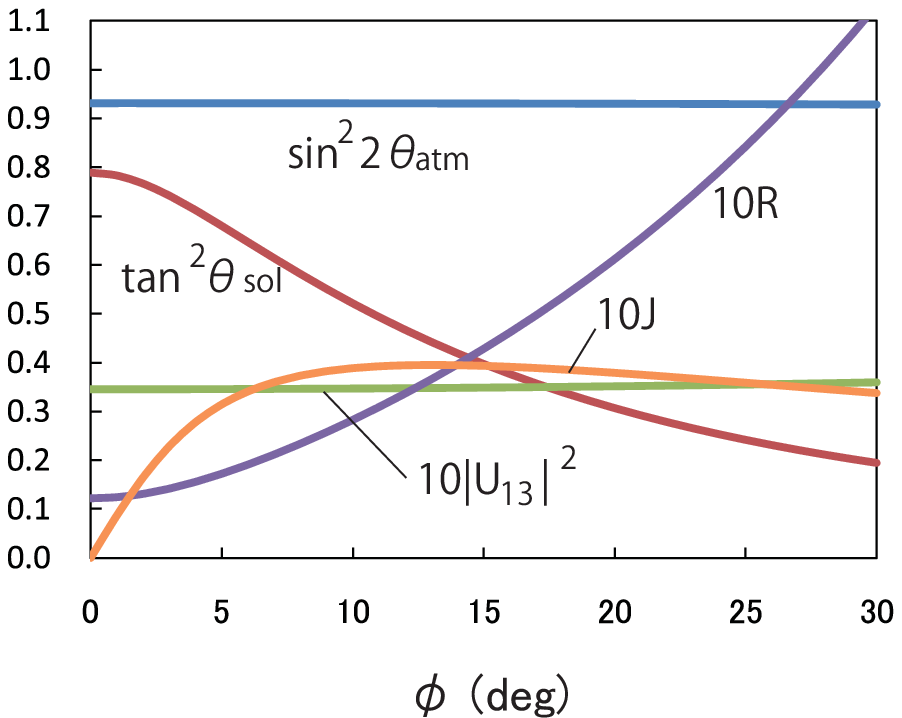}} }

\hspace*{3cm} {\bf (b)}

\begin{quotation}
{\bf Fig.~4}  
Behavior of $J$,  $\sin^2 2\theta_{atm}$, 
$\tan^2 \theta_{sol}$, $R\times 10$, and $|U_{13}|^2\times 10$,  
versus the $CP$ violating phase parameter $\phi$
under the maximal 2-3 symmetry violation $\chi=\pi/2$. 
Figures (a) and (b) correspond to Case A with $a c <0$ and
Case B with $a c >0$, respectively, and those are 
illustrated by taking input values $(a,c)=(-0.255, 0.95)$
and $(a,c)=(+0.27, 0.95)$, respectively.

\end{quotation}

\newpage

{\scalebox{1.0}{\includegraphics{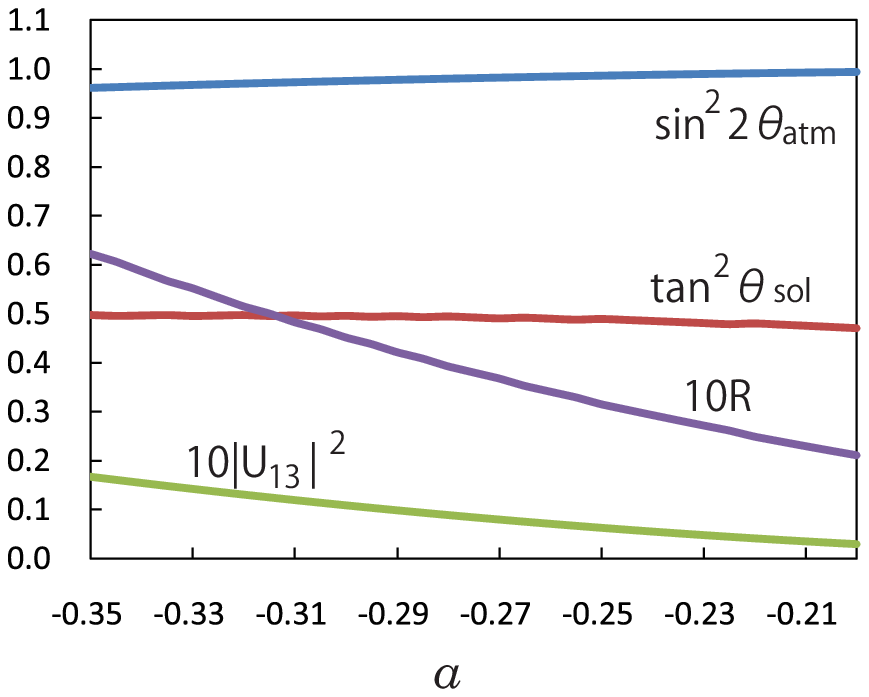}} }

\hspace*{3cm} {\bf (a)}

{\scalebox{1.0}{\includegraphics{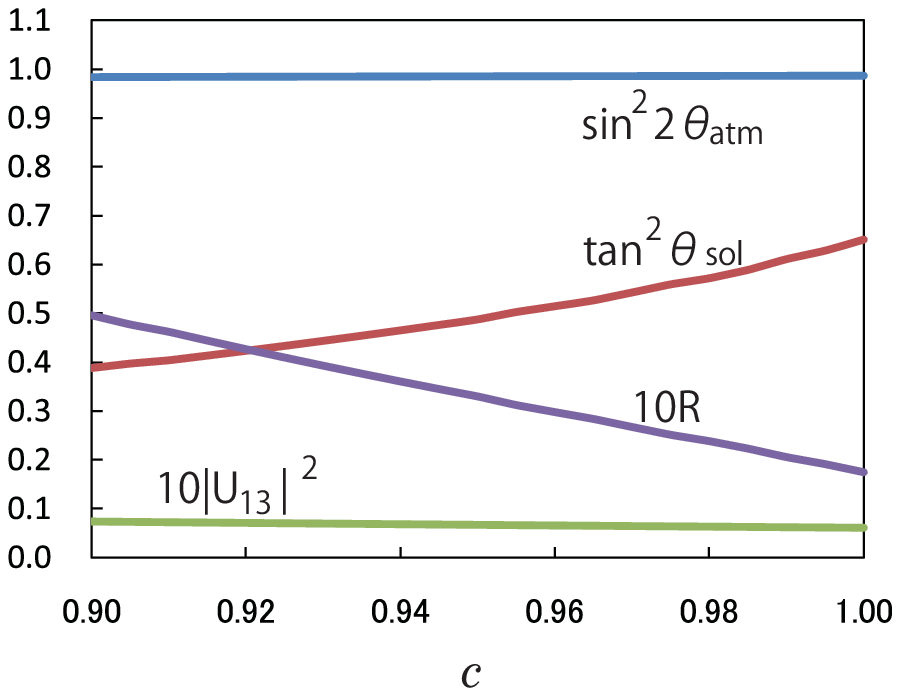}} }

\hspace*{3cm} {\bf (b)}

\begin{quotation}
{\bf Fig.~5}  
$a$-dependence (Fig.~(a)) and $c$-dependence (Fig.~(b)) of the 
predicted values for $J$, $\sin^2 2\theta_{atm}$, 
$\tan^2 \theta_{sol}$, $R\times 10$, and $|U_{13}|^2 \times 10$ 
under the maximal $CP$ violating hypothesis:
the value of $\phi$ is taken as $J$ is maximal for each parameter 
set of $a$ and $c$.
Here, the phase parameter $\chi$ is fixed at $\chi=90^\circ$
at which $J$ takes the maximal value, and the parameter
$c$ is taken as $c=0.95$ for Fig.~(a) and  $a$ as $a=-0.255$ 
for Fig.~(b).
\end{quotation}

\newpage

{\scalebox{0.7}{\includegraphics{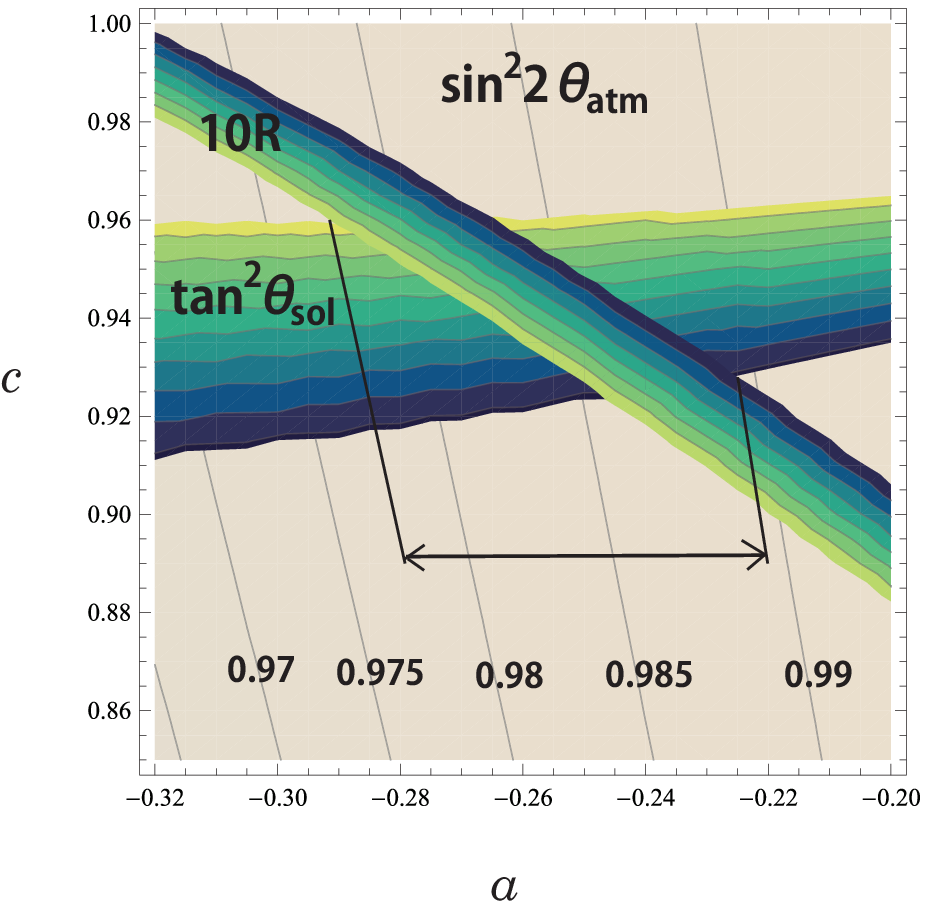}} }
{\scalebox{0.7}{\includegraphics{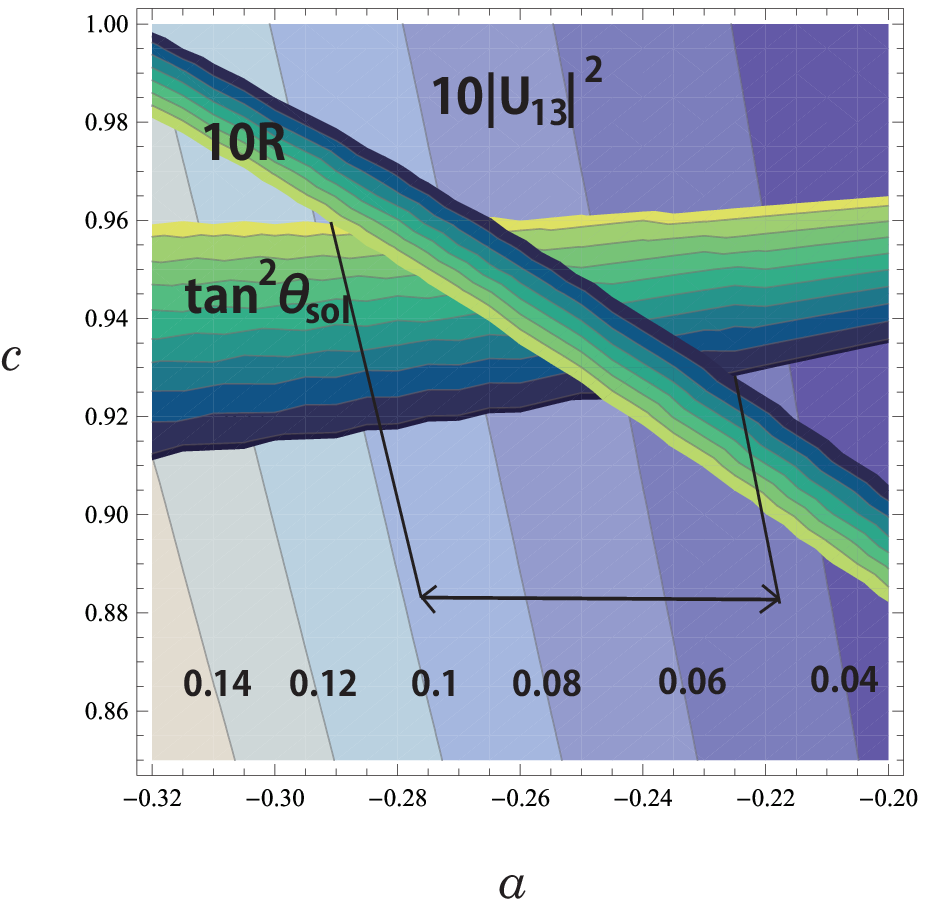}} }

\hspace*{3cm} {\bf (a)}
\hspace*{6cm} {\bf (b)}

{\scalebox{0.7}{\includegraphics{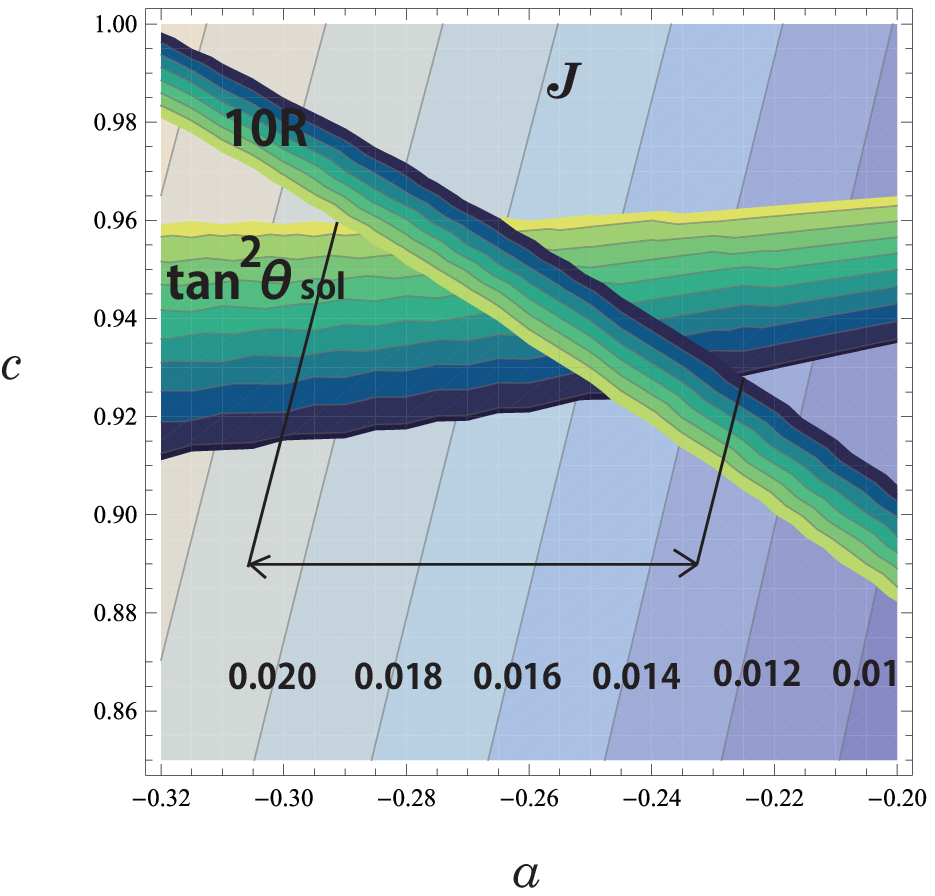}} }

\hspace*{3cm} {\bf (c)}

\begin{quotation}
{\bf Fig.~6}  
Contour plots of predicted values of $\sin^2 2\theta_{atm}$  (Fig.~(a)) , $|U_{13}|^2 \times 10$ (Fig.~(b)), and $J$ (Fig.~(c)) in the $a$-$c$ 
parameter plane under the maximal $CP$ violating hypothesis:
the value of $\phi$ is taken as $J$ is maximal for each parameter 
set of $a$ and $c$. 
Here, the phase parameter $\chi$ is fixed at $\chi=90^\circ$
at which $J$ takes the maximal value.
We also present contour plots of $R\times 10$ and 
$\tan^2 \theta_{sol}$ which are correspond to the experimental 
constraints, $0.0321 \leq |R| \leq 0.0390$ and
$0.428 \leq \tan^2 \theta_{sol} \leq 0.516$, respectively. 
Here the darker region corresponds to smaller contour values.
Note that the overlapped region of the contours of $10R$ and 
$\tan^2\theta_{sol}$ is the allowed region of the $a$-$c$ 
parameter plane of the model.
\end{quotation}

\newpage

{\scalebox{0.7}{\includegraphics{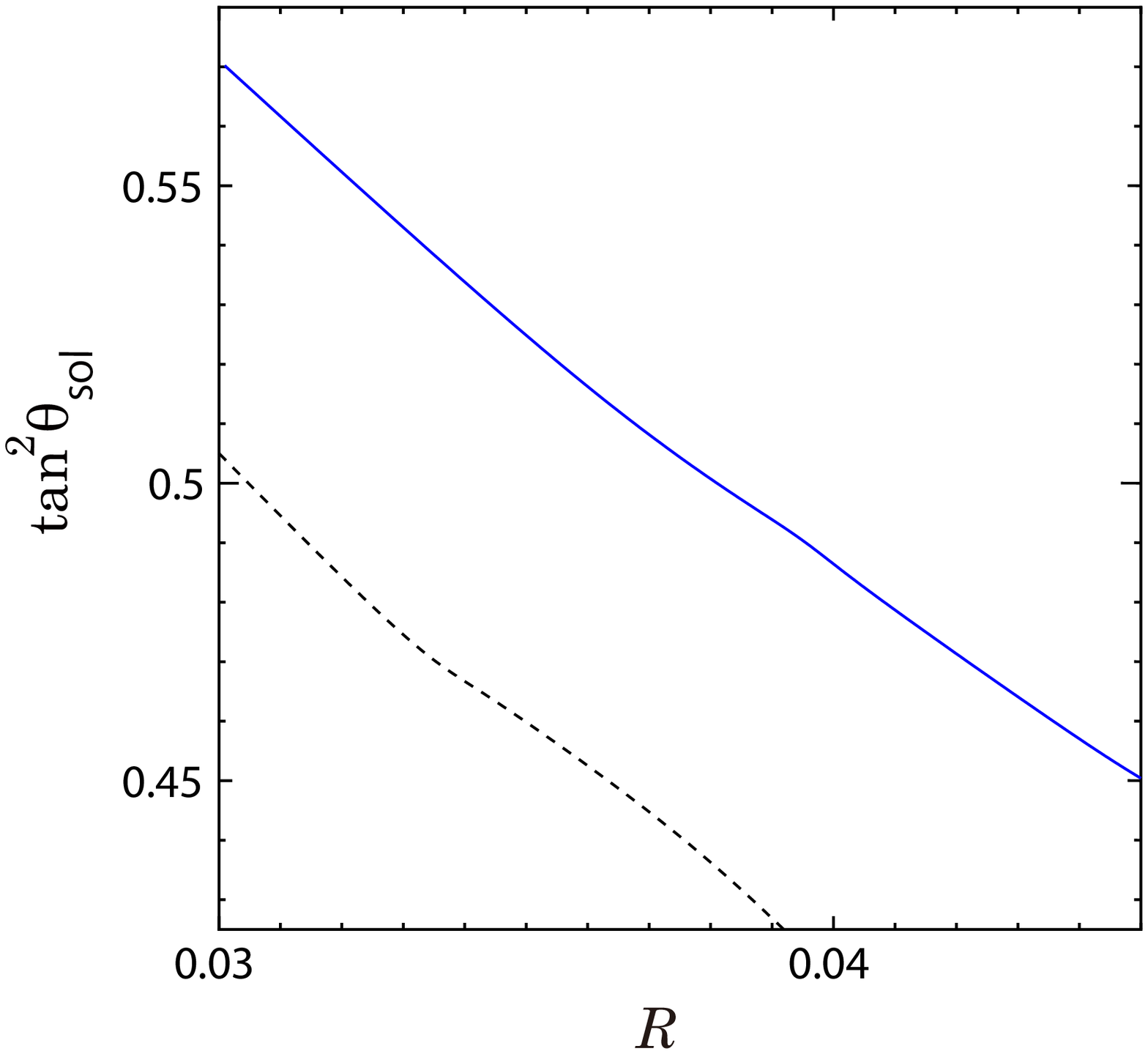}} }

\begin{quotation}
{\bf Fig.~7}  
Behavior of $\tan^2 \theta_{sol}$ vs $R$ under the maximal $CP$ 
violation hypothesis with taking typical values of
$\sin^2 2\theta_{atm}$. 
The solid and dashed curves correspond to the cases with $\sin^2 2\theta_{atm}=0.980$ 
and  $\sin^2 2\theta_{atm}=0.985$, respectively.
The predicted values of 
$|U_{13}|^2$ for $\sin^2 2\theta_{atm}=0.980$
and $\sin^2 2\theta_{atm}=0.985$ are $|U_{13}|^2 \simeq 0.0087$ 
and $|U_{13}|^2 \simeq 0.0066$, respectively, in this parameter
region of $(a,c)$.
\end{quotation}

\end{document}